\documentclass[a4paper,showkeys,
amsmath,amssymb,superscriptaddress,pra,twocolumn,floatfix,showpacs]{revtex4}
\usepackage[utf8]{inputenc}
\usepackage{amsmath,amssymb}
\usepackage{amsthm}
\usepackage{graphicx}
\usepackage{hyperref}

\newtheorem{theorem}{Theorem}

\newcommand{\ket}[1]{| #1
	\rangle} \newcommand{\bra}[1]{\langle #1 |} \newcommand{\ketbra}[2]{| #1 
	\rangle
	\langle #2 |} \newcommand{\braket}[2]{\bra{#2}{#1}\rangle}
 \newcommand{\kr}{\otimes}
\newcommand{\tr}{\mathrm{Tr}}

\newcommand{\e}{\mathbb{e}}

  \newcommand{\1}{{\rm
		1\hspace{-0.9mm}l}}

\begin{document}
\author{Dariusz Kurzyk}
\affiliation{Institute of
Theoretical and Applied Informatics, Polish Academy of Sciences, Ba{\l}tycka 5,
44-100 Gliwice, Poland}

\author{{\L}ukasz Pawela\footnote{Corresponding author, E-mail:
lpawela@iitis.pl}}
\affiliation{Institute of
Theoretical and Applied Informatics, Polish Academy of Sciences, Ba{\l}tycka 5,
44-100 Gliwice, Poland}

\author{Zbigniew Pucha{\l}a}
\affiliation{Institute of Theoretical and Applied Informatics, Polish Academy
of Sciences, Ba{\l}tycka 5, 44-100 Gliwice, Poland}
\affiliation{Institute of Physics, Jagiellonian University, ul. Stanis\l{}awa 
\L{}ojasiewicza 11, 30-348 Krak\'o{}w, Poland}

\title{Unconditional Security of a $K$-State Quantum Key Distribution Protocol} 
\date{February 14, 2018}

\begin{abstract} 
Quantum key distribution protocols constitute an important part of quantum
cryptography, where the security of sensitive information arises from the laws
of physics. In this paper we introduce a new family of key distribution
protocols and we compare its key
with the well-known protocols such as BB84, PBC0 and
generation rate to the well-known protocols such as BB84, PBC0 and
R04. We also state the security analysis of these protocols based on the
entanglement distillation and CSS codes techniques.
	

\end{abstract}

\maketitle
		
\section{Introduction}

Quantum cryptography is a blooming field of scientific research, where quantum
phenomena are applied to securing sensitive information. Usually, cryptographic
systems are based on the key distribution mechanisms and security of the systems
depend on computational complexity. The security of quantum cryptography arises
from the laws of quantum physics. Thus, quantum cryptography does not impose
limitation on the eavesdropper's technology, which is limited by the laws that no
one can ever overcome. Scenarios of quantum key distribution (QKD) protocols are
based on the assumption that secret key is shared by Alice and Bob, and only a
small amount of information can be leaked to an eavesdropper Eve. The first QKD
protocol, BB84~\cite{bennett1984update}, became a motivation for expanding 
research in this area. As a consequence, Mayers in \cite{mayers1996quantum}
proved the unconditional security of this protocol on a noisy channel against a
general attack. Quantum entanglement and the violation of Bell's theorem was 
introduced to the BB84 protocol by Ekert \cite{ekert1991quantum}. Next,
Bennett proposed a simple protocol B92 \cite{bennett1992quantum} based on two
nonorthogonal states. Unconditional security analysis of this protocol was
performed by Tamaki \emph{et al.} in
\cite{tamaki2003unconditionally,tamaki2004unconditional} and by Quan \emph{et
al.} in \cite{quan2002simple}. Generalization of BB84
protocol using conjugate bases, i.e. six states was discussed by Bru{\ss}\cite{bruss1998optimal}.
Subsequently, Phoenix \emph{et al.} \cite{phoenix2000three} introduced PBC00
protocol and they showed that key bits can be generated more efficiently by the
usage of three mutually non-orthogonal states. Renes developed the key creation
protocols R04 \cite{renes2004spherical} for two qubit-based spherical codes,
which is a modified version of the PBC00 protocol. The R04 protocol allows one
to use all conclusive events for key extraction. In
\cite{boileau2005unconditional}, Boileau \emph{et al.} proved the unconditional
security of the trine spherical code QKD protocol, which concerns also to PBC00
and R04 protocols. The experimental realization of PBC00 and R04 protocols were
proposed in \cite{senekane2015six} and \cite{schiavon2016experimental}. New
results referring to asymptotic analysis of three-state protocol can be found in
\cite{krawec2016asymptotic}. Scarani \emph{et al.} \cite{scarani2004quantum}
introduced a protocol SARG04, which is resistant to photon number splitting attacks.
QKD protocols have become a subject of profound research and various
unconditional security analysis of QKD protocol can be found in
\cite{mayers2001unconditional,koashi2003secure,renner2005information,biham2006proof,inamori2007unconditional}, while
 for a review see \cite{scarani2009security}.

In this paper we propose a new class of QKD protocols with security analysis
performed by the use of techniques similar as in
\cite{tamaki2003unconditionally,boileau2005unconditional}. It means that the
proposed protocol was considered as entanglement distillation protocol (EDP)
\cite{bennett1996mixed,lo1999unconditional}. Subsequently, similarly as in case
of BB84 \cite{shor2000simple}, CSS codes
\cite{calderbank1996good,steane1996multiple} were used to the security proof.

\section{$K$-state protocol}

Let us introduce a class of QKD protocols, which can be reducible to the
well-known PBC00 protocol~\cite{phoenix2000three}. Assume that Alice and Bob
would like to share $N$ secret bits $b_i$. Then, the protocol is as follows.

\subsection*{\textbf{Protocol 1 (P1)}}
\begin{enumerate}
\item Alice and Bob share $N$ pairs of maximally entangled two-qubit states
$\rho_{AB}=\ket{\phi^-}\bra{\phi^-}$, where $\ket{\phi^-} =
\frac{1}{\sqrt{2}}(\ket{01}-\ket{10})$.

\item She chooses $K$ states $\ket{\psi_k}=\cos(a+\theta_k)\ket{0} +
\sin(a+\theta_k)\ket{1}$, where $a\in[0,2\pi)$ is a constant and
$\theta_k=\frac{2 k \pi}{K}$ for $k\in\{0,...,K-1\}$. The states $\ket{\psi_k}$
are grouped into pairs $S_k=\{\ket{\psi_k},\ket{\psi_{k+1\mod K}}\}$. \item
Subsequently, Alice measures her parts of the states $\rho_{AB}$ using the POVM
$\{\frac2K\ketbra{\psi_k^\perp}{\psi_k^\perp}\}_k$, where $\ket{\psi_k^\perp}$
is orthogonal to $\ket{\psi_k}$. Detection of the state $\ket{\psi_k^\perp}$
after measurement is equivalent to sending a state $\ket{\psi_k}$ to Bob.

\item For each bit $b_i$ to be encoded, Alice chooses at random $r_i\in{0,\ldots,k-1}$. 
The choice of $r_i$ determines the encoding base $S_{r_i}$. Thus, the bit value 
is equal to $b_i=k-r_i\mod K$.

\item Alice publicly announces when all of her measurements are done. 

\item Bob prepares measurements described by the POVM
$\{\frac2K\ketbra{\psi_k^\perp}{\psi_k^\perp}\}_k$ and announces when the measurements are done. Next, Alice
sends sequences of values $r_i$ to Bob. If the $i$-th Bob's measurement
outcome is $\ket{\psi_k^\perp}$, then Bob decodes $b_i=0$ (if $k=r_i$) or
$b_i=1$ (if $k=r_i-1 \mod K$). In other cases, the events are regarded as
inconclusive. These results are discarded.

\item Half of randomly chosen conclusive events are used in the estimation of a
bit error rate. If the bit error rate is to high, then they abort the protocol.

\item In the end, they use a classical error correction and privacy
amplification protocols.
\end{enumerate}	

Notice that for $K=3$ and an appropriate choice of $a$, the above scenario is
equivalent to the PBC00 protocol \cite{phoenix2000three}. It can also be shown 
that protocols of this class achieve the highest key rate for $K=3$.

\section{POVM enhancement}

Now we consider a modification of the above protocol. Steps 1--4 are the same 
as in the previous protocol.

\subsection*{\textbf{Protocol 2 (P2)}}
\begin{enumerate}
\item[5$^\prime$.] Alice publicly announces when all her measurements are 
done and she sends sequences of values $r_i$ to Bob. 

\item[6$^\prime$.] For each $r_i$, Bob prepares an unambiguous measurement 
described by the POVM
\begin{equation}\label{eq:new_povm}
\begin{split}
\{&\Pi_{r_i-1},\Pi_{r_i},\Pi_{fill}\}=\\
\Big\{
& 
\frac{1}{\lambda}\ketbra{{\psi}^\perp_{r_i}}{{\psi}^\perp_{r_i}},\frac{1}{\lambda}\ketbra{{\psi}^\perp_{r_i+1\!\!\!\!\mod 
		K}}{{\psi}^\perp_{r_i+1\!\!\!\!\mod K}}\\ 
&\1_2 - \frac{1}{\lambda} \big(\ketbra{{\psi}^\perp_{{r_i}}}{{\psi}^\perp_{{r_i}}}+\ketbra{{\psi}^\perp_{{r_i}+1\!\!\!\!\mod 
		K}}{{\psi}^\perp_{{r_i}+1\!\!\!\!\mod 
		K}}\big)\Big\} 
\end{split}
\end{equation}
where $\lambda = \frac{\sqrt{2}}{2} \sqrt{\cos\frac{4\pi}{K}+1} + 1 = 1 
+|\cos(\frac{2\pi}{K})|$. The value $\lambda$ is determined as a maximal 
eigenvalue of 
\begin{equation}
\ketbra{\tilde{\psi}_{r_i}}{\tilde{\psi}_{r_i}} + 
\ketbra{\tilde{\psi}_{r_i+1\!\!\!\!\mod K}}{\tilde{\psi}_{r_i+1\!\!\!\!\mod K}}.
\end{equation}
If the $i$-th Bob's measurement outcome is $\ket{\psi_k^\perp}$, then Bob
decodes $b_i=0$ (if $k=r_i$) or $b_i=1$ (if $k=r_i-1 \mod K$). In other cases,
the events are regarded as inconclusive. Finally, Alice and Bob discard results
of measurement, which are inconclusive.
\end{enumerate}	
Steps 7. and 8. are again the same as in the previous protocol. For $K=3$, 
the characteristics of the protocols P1 and P2 are the same and are equivalent 
to the characteristics of the PBC00 protocol.

\section{Security analysis}

Similarly to
\cite{tamaki2003unconditionally,boileau2005unconditional,shor2000simple}, we
consider an entanglement distillation protocol (EDP), which can be reduced to a
QKD protocol equivalent to the above scheme.  Firstly, we transform the vectors
$\ket{\psi_i}$ by the rotation operator $R(-\eta)$, where
$R_y(\theta)=\left(\begin{matrix} \cos \theta & -\sin \theta\\ \sin\theta & 
\cos\theta  \end{matrix}\right)$ and $\eta=\arccos(\braket{\psi_0}{\psi_1})/2 + 
\arctan\left(\frac{\braket{1}{\psi_0}}{\braket{0}{\psi_0}}\right)$. After this
transformation, we get states $\ket{\tilde{\psi_i}}=R(-\eta)\ket{{\psi_i}}$, where
$\ket{\tilde{\psi_0}}=\cos(\frac{\pi}{K})\ket{0} + \sin(\frac{\pi}{K})\ket{1}$
and $\ket{\tilde{\psi_1}}=\cos(-\frac{\pi}{K})\ket{0} +
\sin(-\frac{\pi}{K})\ket{1}$. This transformation has no impact on the
protocol, but is important in the security analysis. Assume that Alice prepares
many pairs of qubits in the entangled state
$\ket{\psi}=\frac{1}{\sqrt{2}}(\ket{+}\ket{\tilde{\psi}_0}+\ket{-}\ket{\tilde{\psi}_1})$,
 where $\ket{\pm}=\frac{1}{\sqrt{2}}(\ket{0}\pm\ket{1})$ and the basis $\{\ket{+},\ket{-}\}$ 
 will be denoted by $\pm$-basis. Next, she 
randomly chooses a string of trit values $r_i$ and applies $R_y(\theta_{r_i})$ 
on the second qubit of every pair. After that, she sends qubits to Bob through 
a quantum channel. Alice announces the values of $r_i$. Next, Bob performs 
\textit{local filtering operations} 
\cite{gisin1996hidden,bennett1996concentrating,horodecki1997inseparable} 
$F=\frac{1}{\sqrt{2\lambda}}\Big((\ket{\tilde{\psi}^\perp_0} +
\ket{\tilde{\psi}^\perp_1})\bra{0} + (\ket{\tilde{\psi}^\perp_0} - 
\ket{\tilde{\psi}^\perp_1})\bra{1} \Big)$ and operation $R_y(-\theta_{r_i})$ on the received qubits. Next, half of the states are
used to determine the number of bit errors after application of $\pm$-basis 
measurements by Alice and Bob. If the number of error is too high, then the 
protocol is aborted. Remaining qubits are used to distill Bell states by 
an EDP based on CSS codes. Alice and Bob perform $\pm$-basis measurements 
on Bell states to obtain a secret key.

Notice that $R_y(\theta_{r_i})\ket{\tilde{\psi}_j}=\ket{\tilde{\psi}_{r_i+j
\mod K}}$ and Alice's operation related to measurement
$\{\frac{1}{\lambda}\ketbra{\tilde{\psi}_i^\perp}{\tilde{\psi}_i^\perp}\}_i$ on her state
are equivalent to $\pm$-basis measurement on the state $\big(\1_2\kr
R_y(\theta_{r_i})\big)\ket{\psi}$.  The filtering operations $F$, rotation operation 
$R_y(-\theta_{r_i})$ and $\pm$-basis measurement performed by 
Bob can by described by the following POVM
\begin{equation}\label{eq:edp_povm}
\begin{split}
&\quad\quad \{R_y(\theta_{r_i}) F^\dagger \ketbra{+}{+} F  
R_y(\theta_{r_i})^\dagger,\\
&\quad\quad R_y(\theta_{r_i}) F^\dagger \ketbra{-}{-} F  
R_y(\theta_{r_i})^\dagger,\\
&\quad\quad R_y(\theta_{r_i}) (\1_2 - F^\dagger F)  R_y(\theta_{r_i})^\dagger\}.\\
\end{split}
\end{equation}
This measurement is equivalent to the POVM
\begin{equation}
\begin{split}
\{&\tilde{\Pi}_{r_i-1},\tilde{\Pi}_{r_i},\tilde{\Pi}_{fill}\}=\\
\Big\{
& 
\frac{1}{\lambda}\ketbra{\tilde{\psi}^\perp_{r_i}}{\tilde{\psi}^\perp_{r_i}},\frac{1}{\lambda}\ketbra{\tilde{\psi}^\perp_{r_i+1\!\!\!\!\mod 
		K}}{\tilde{\psi}^\perp_{r_i+1\!\!\!\!\mod K}},\\ 
&\1_2 - \frac{1}{\lambda} (\ketbra{\tilde{\psi}^\perp_{{r_i}}}{\tilde{\psi}^\perp_{{r_i}}}+\ketbra{\tilde{\psi}^\perp_{{r_i}+1\!\!\!\!\mod 
		K}}{\tilde{\psi}^\perp_{{r_i}+1\!\!\!\!\mod 
		K}})\Big\}.
\end{split}
\end{equation}

In \cite{shor2000simple}, Shor and Preskil have shown that if the bound of
estimations of bit and phase error decreases exponentially as $N$ increases,
then Eve's information on secret key is exponentially small. This approach was
used to prove the unconditional security of the Bennet 1992 protocol, by Tamaki
\emph{et al.} \cite{tamaki2003unconditionally}, and the PBC00 and R04 protocols,
by Boileau \emph{et al.} \cite{boileau2005unconditional}. These proofs were
based on the usage of reduction to an entanglement distillation protocol
initialed by a local filtering process. Subsequently, we will prove the security
of the above entanglement distillation protocol in the same manner as in
\cite{tamaki2003unconditionally,boileau2005unconditional,shor2000simple}.

Assume that $\{p_{b}^{(i)}\}_{i=1}^N$ and $\{p_p^{(i)}\}_{i=1}^N$ are sets of
probabilities of a bit error and a phase error respectively on the $i$-th pair
after Alice and Bob have done the same measurements on $i-1$ previous pairs.
Thus $p_b^{(i)}$ and $p_p^{(i)}$ depend on previous results. Moreover, we
introduce $e_b$ and $e_p$ as rates of bit error and phase error in all
conclusive results respectively. Estimations of bit and phase error rates will
be performed by to use of Azuma's inequality \cite{azuma1967weighted} as in
\cite{boileau2005unconditional}.

\begin{theorem}[\cite{azuma1967weighted}]
Let $\{X_i:i=0,1,\dots\}$ be a martingale sequence and for each $k$ it holds
that $|X_k-X_{k-1}|\leq c_k$. Then for all integers $N\geq 0$ and real numbers
$\gamma\geq0$
\begin{equation}
P(|X_N-X_{0}| \geq \gamma) \leq 2 \e^{-\frac{\gamma^2}{2 \sum_{k=1}^N c_k^2}}.
\end{equation}\label{azuma}
\end{theorem}
Notice that for $c_k=1$ we get	
\begin{equation}
P(|X_N-X_{0}| \geq \gamma) \leq 2 \e^{-\frac{\gamma^2}{2 N}}.
\end{equation}

As a result of the Azuma's inequality, $Ce_b$ is exponentially close to $e_p$
($Ce_b=e_p$) for particular constant $C$, if $Cp^{(i)}_b=p_p^{(i)}$ is satisfied
for all $i$. Assume that Eve can perform any coherent attack $E^{(i)}$ on
qubits sent by Alice such that  $\sum_i E^{(i)\dagger} E^{(i)}\leq \1$. The
general equation for the $i$\textsuperscript{th} state can be described by a
mixed state 
\begin{equation}
\rho^{(i)} = \frac1K\sum_{k=0}^K
\ketbra{\phi_k^{(i)}}{\phi_k^{(i)}},
\end{equation}
where
\begin{equation}
\ket{\phi_k^{(i)}}=\1_A\kr\big(FR(-\theta_k)E^{(i)}R(\theta_k)\big)\ket{\psi}.
\end{equation}
Let us introduce the notation
$\ket{\Phi^\pm}=\frac{1}{\sqrt{2}}(\ket{+}\ket{+}\pm\ket{-}\ket{-})$ and
$\ket{\Psi^\pm}=\frac{1}{\sqrt{2}}(\ket{+}\ket{-}\pm\ket{-}\ket{+})$.
Since the probability of sharing by Alice and Bob a Bell state $\ket{\Phi^+}$
is equal to the probabilities of a bit error $p_{b}^{(i)}$ and phase error
$p_{p}^{(i)}$ on the $i$-{th} respectively, thus
\begin{equation}
\begin{split}
p_{b}^{(i)}&=\frac{1}{Z^{(i)}}\big(\bra{\Psi^+} \rho^{(i)}\ket{\Psi^+} + 
\bra{\Psi^-_Z} \rho^{(i)}\ket{\Psi^-} \big)\\
p_{p}^{(i)}&=\frac{1}{Z^{(i)}}\big(\bra{\Phi^-} \rho^{(i)}\ket{\Phi^-} + 
\bra{\Psi^-} \rho^{(i)}\ket{\Psi^-} \big),
\end{split}
\end{equation}
where
\begin{equation}
\begin{split}
Z^{(i)}&=\big(\bra{\Psi^+} \rho^{(i)}\ket{\Psi^+} + \bra{\Psi^-} 
\rho^{(i)}\ket{\Psi^-}\\ &+ \bra{\Phi^+} \rho^{(i)}\ket{\Phi^+} + 
\bra{\Phi^-} \rho^{(i)}\ket{\Phi^-} \big).
\end{split}
\end{equation}

It can be checked that
\begin{equation}\label{eq:C}
C = \frac{p_{p}^{(i)}}{p_{b}^{(i)}} =  1+|\braket{\psi_0}{\psi_1}|^2 = 1 +
\cos^2\Big(\frac{2\pi}{K}\Big).
\end{equation}

Similarly as in \cite{boileau2005unconditional}, we calculate the key rate $S$
from the following formula
\begin{equation}\label{eq:key_rate}
S=p_{c}(e_{b})\big(1-h(e_{b})-h(e_{p})\big),
\end{equation}
where $h(x)=-x\log_2 x -(1-x)\log_2(1-x)$ and $p_c(e_{b})$ is the probability of
a conclusive result. Since $Ce_b=e_p$, we get
\begin{equation}\label{eq:key}
S=p_{c}(e_{b})\big(1-h(e_{b})-h(Ce_{b})\big).
\end{equation}
Notice that for a bit value $b=0$ we get outcome probabilities

$\{0,\frac1\lambda 
|\braket{\tilde{\psi}_{r_i}}{\tilde{\psi}^\perp_{r_i+1\!\!\!\! \mod K}}|^2,  
1-\frac{1}{\lambda}|\braket{\tilde{\psi}_{r_i}}{\tilde{\psi}^\perp_{r_i+1\!\!\!\!
 \mod K}}|^2\}$
and for $b=1$ we get  

$\{\frac1\lambda 
|\braket{\tilde{\psi}_{r_i+1\!\!\!\! \mod K}}{\tilde{\psi}^\perp_{r_i}}|^2, 0,
1-\frac{1}{\lambda}|\braket{\tilde{\psi}_{r_i+1\!\!\!\! \mod 
K}}{\tilde{\psi}^\perp_{r_i}}|^2\}$. Hence, it can be checked that 
$|\braket{\tilde{\psi}_{r_i}}{\tilde{\psi}^\perp_{r_i+1\!\!\!\! \mod 
K}}|^2=|\braket{\tilde{\psi}_{r_i+1\!\!\!\! \mod 
K}}{\tilde{\psi}^\perp_{r_i}}|^2=\sin^2(\frac{2\pi}{K})$. Thus, the probability 
of a conclusive result, with the assumption that $e_b=0$, is equal to
\begin{equation}\label{eq:init_pc}
p_c(0)=\frac{\sin^2(\frac{2\pi}{K})}{1+|\cos(\frac{2\pi}{K})|},
\end{equation} 
which can be simplified to $p_c(0)=2\sin^2(\frac{\pi}{K})$ for $k>3$.

Generally, $p_c$ can be expressed as
\begin{equation}\label{eq:pc}
p_c(e_{b})=
\frac{\sin^2(\frac{2 \pi}{K})}{\lambda (1-2e_b \cos^2(\frac{2\pi}{K}))},
\end{equation}
which was derived in Appendix~\ref{app:pc_new}. Notice, that for $K=3$, 
Eq.~(\ref{eq:pc}) is reduced to $p_c(e_{b})=\frac{1}{2-e_b}$, which corresponds to 
probability of conclusive results in PBC00 protocol.

\begin{figure}[!b]
\centering
\includegraphics[scale=0.87]{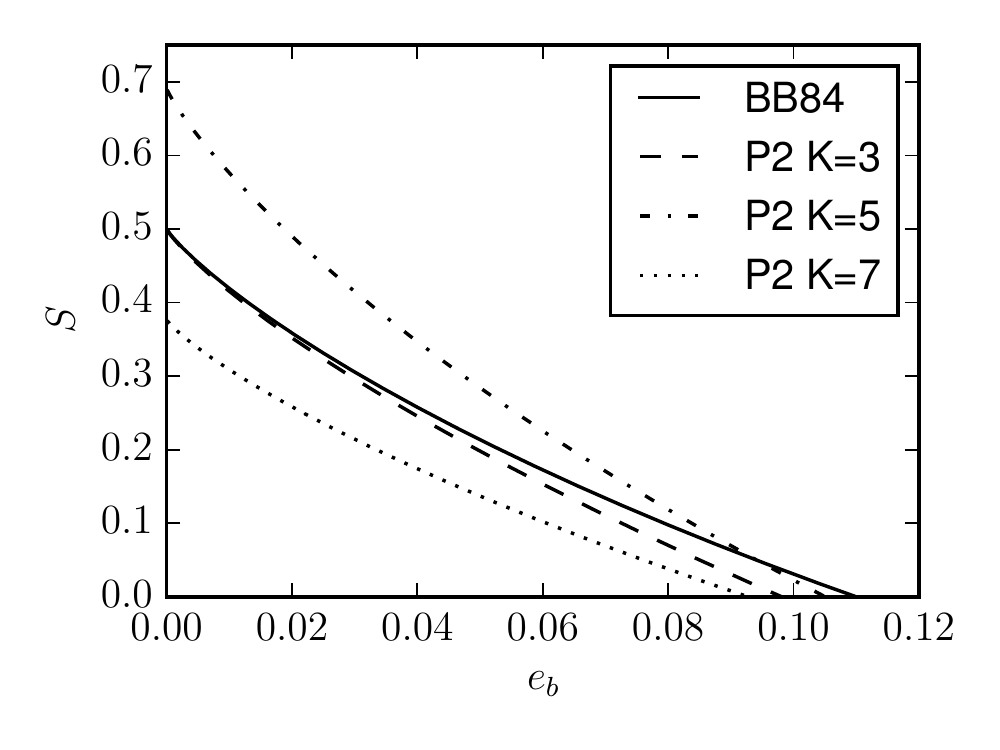}
\caption{Comparison of key rates depending on $e_{b}$ for different setups of 
the P2 protocol. Notice that for $K=5$ we get the best key rates. For $K=7$, 
these drop below the values of obtained for $K=3$. We also show the key rates 
of the BB84 protocol for comparison.}\label{fig:comparison}
\end{figure}

In the case of BB84 protocol, bit error rate is equal to phase error rate.
Thus $C=1$ and $p_c(e_{b})=\frac12$. In the case of $PBC00$, $C=\frac54$ and
$p_c(e_{b})=\frac{1}{2-e_b}$. From Eq.~(\ref{eq:key}) we get that $e_b\approx 11.0\%$
for BB84 protocol and $e_b\approx9.81\%$ for PBC00 protocol. It can be checked 
that an interesting case is for $K=5$, where $C=\frac18(11-\sqrt{5})$ and
$e_b\approx10.5\%$. Comparison of proposed protocol with BB84 and PBC00
protocols is shown in Fig.~\ref{fig:comparison}. As we can see, the best key
ratio is for $K=5$.

\section{Conclusion} In this paper we have introduced a new class of quantum key
distribution protocols. We have also provided unconditional security analysis of
this protocol. We have shown, that there exists $5$-state protocol with 
reasonably high key rate for small bit-flip error rates.

\section*{Acknowledgement} The authors acknowledge the support by the Polish
National Science Center under the Project Number 2015/17/B/ST6/01872.

\bibliography{../bib} 
\bibliographystyle{unsrt}

\appendix
\section*{Appendix}

\section{Probability of conclusive events}\label{app:pc_new} Notice that if Alice performs
$\ket{\psi^\perp_i}$, where $\ket{\psi_i}\in S_r$, $i=r$ or $i=r+b\mod K$ and
Bob chooses $\ket{\psi_i^\perp}$, then it corresponds to an error. In the case
when Bob chooses a state which corresponds to $S_r$ but is not orthogonal to
Alice's state, then Bob can correctly conclude the state $\ket{\psi_i}$.

Let $n_g, n_e, n_i$ denote the numbers of good conclusive, error conclusive
and inconclusive events respectively. Beside that, let $n_t = n_g+n_e+n_i$
and thus $1=\frac{n_g}{n_t}+\frac{n_e}{n_t}+\frac{n_i}{n_t}$. Assume that after
Alice sent $r$ to Bob, Bob performs measurement
described by POVM
\begin{equation}
\begin{split}
\{&\Pi_{r},\Pi_{r+1},\Pi_{fill}\}=\\
\Big\{
& 
\frac{1}{\lambda}\ketbra{{\psi}^\perp_r}{{\psi}^\perp_r},\frac{1}{\lambda}\ketbra{{\psi}^\perp_{r+1\!\!\!\!\mod 
		K}}{{\psi}^\perp_{r+1\!\!\!\!\mod K}},\\ 
&\1_2 - \frac{1}{\lambda} (\ketbra{{\psi}^\perp_{r}}{{\psi}^\perp_{r}}+\ketbra{{\psi}^\perp_{r+1\!\!\!\!\mod 
		K}}{{\psi}^\perp_{r+1\!\!\!\!\mod 
		K}})\Big\}.
\end{split}
\end{equation}

Now, we suppose that $b=0$ and Eve simulates a noisy channel, where state
$\ketbra{{\psi}_r}{{\psi}_r}$ evolves as
$\rho_B=(1-p)\ketbra{{\psi}_r}{{\psi}_r}+\frac{p}{2}\1_2$. Next, Bob performs
measurement and receives measurement outcomes with probabilities $\{\tr \Pi_r \rho_B =
\frac{p}{2\lambda}, \tr
\Pi_{r+1} \rho_B = \frac{p}{2\lambda} +
\frac{1-p}{\lambda}\sin^2(\frac{2\pi}{K}), \tr \Pi_{fill} \rho_b = 1-
\frac{p}{\lambda}-\frac{1-p}{\lambda}\sin^2(\frac{2\pi}{K}\}$.

A bit error rate $e_b$ is defined as the rate of error in conclusive results. Hence 
\begin{equation}
e_b = \frac{n_e}{n_e+n_g} \quad \text{and} \quad n_e=\frac{e_b}{1-e_b} n_g.
\end{equation}
Notice that error $e_b$ can be estimated as
\begin{equation}
e_b = \frac{\tr \Pi_{r+1} \rho_B}{\tr \Pi_{r+1} \rho_B+ \Pi_{fill} 
\rho_b}=\frac{p}{ 2(1-p)\sin^2(\frac{2\pi}{K}) + 2p}.
\end{equation}

Now, let us determine a ratio
\begin{equation}
\begin{split}
D &= \frac{\tr \Pi_{r+1} \rho_B}{\tr \Pi_{fill} \rho_B} = 
\frac{2(1-p)\sin^2(\frac{2\pi}{K})+p}{2\lambda - 
2(1-p)\sin^2(\frac{2\pi}{K})-2p}\\
&=\frac{2(1-e_b)  \sin^2(\frac{2 \pi}{K})}{2\lambda 
(1-2e_b\cos^2(\frac{2\pi}{K}))-2\sin^2(\frac{2\pi}{K})}.
\end{split}
\end{equation}

From the central limit theorem and the above calculation we get
\begin{equation}
\frac{n_g}{n_t}=\frac{D n_i}{n_t}+O(\epsilon).
\end{equation}
Continuing we obtain
\begin{equation}
\begin{split}\label{eq:pc_calc_new}
1&=\frac{n_g}{n_t}+\frac{n_e}{n_t}+\frac{n_i}{n_t}=\frac{n_g}{n_t}+\frac{e_b}{1-e_b}\frac{n_g}{n_t}+\frac{n_i}{n_t}\\
&\approx\frac{Dn_i}{n_t}+\frac{e_b}{1-e_b}\frac{Dn_i}{n_t}+\frac{n_i}{n_t}\\	
&\approx\frac{D+1-e_b}{1-e_b} \frac{n_i}{n_t}
\end{split}
\end{equation}
and
\begin{equation}\label{eq:pc_calc_2}
p_c=1-\frac{n_i}{n_t} = \frac{D}{D+1-e_b}=
\frac{\sin^2(\frac{2 \pi}{K})}{\lambda (1-2e_b \cos^2(\frac{2\pi}{K}))}.
\end{equation}

\end{document}